\documentclass[A4paper, twocolumn]{article}
\usepackage{graphicx}
\usepackage{subfigure}
\usepackage{amsfonts,amssymb}
\usepackage{amsmath}
\usepackage{bbold}
\usepackage{dsfont}
\usepackage{color}
\usepackage{multicol}
\usepackage{abstract}

\newtheorem{proposition}{Proposition}
\newtheorem{corollary}{Corollary}

\begin{document}

\title{Quantifying nonclassicality and entanglement of Gaussian states}

\author{Xiao-yu Chen \\ 
{\small {School of Information and Electrical Engineering, Hangzhou City University, Hangzhou {\rm 310015}, China }}}

\date{}

\twocolumn[
\maketitle

\begin{onecolabstract}

Quantification of nonclassicality and entanglement in a quantum state is crucial for quantum advantage in information processing and computation. Robustness is one of the tractable measures for quantifying quantum resources. Gaussian states are important in continuous variable quantum information for their theoretically simple and experimentally easily accessible. We provide the method of how to calculate the robustness of nonclassicality and enatnglement for Gaussian states.  The robustness of nonclassicality or entanglement is demonstrated analytically for one-mode, two-mode Gaussain states and multimode symmetric Gaussian states, the result shows a clear physical meaning for the origin of nonclassicality and entanglement. For squeezed thermal states, the nonclassicality is equal to the entanglement for the two-mode case, while they are far apart for multimode cases.
   
\end{onecolabstract}
]



In continuous variable systems, one of the main tasks is to quantify the entanglement and (optical) nonclassicality of a given quantum state. The quantification of nonclassicality was only proposed recently, although optical nonclassicality has been discussed for decades. The entanglement measures have been introduced for years. Basically, there are two kind of measures, one is the kind with clear physical meaning but is very difficult to be calculated such as entanglement of formation and distillable entanglement\cite{Bennett}. The other is defined with some distance thus is relatively easy to be evaluated but without a clear physical meaning such as relative entropy of entanglement \cite{Vedral} and geometric measure. A measure with both merits is the robustness, it is related to the quantum advantage of subchannel discrimination\cite{Takagi}\cite{Piani}, thus has an operational meaning.    

Robustness of nonclassicality ( entanglement) is the minimal amount of other quantum state which can be added to the resourceful quantum state such that the resultant state becomes classical (separable)\cite{LamiPRA}\cite{Vidal}\cite{Steiner}. Nonclassicality can be seen as a special kind of quantum coherence, the robustness of coherence has been developed\cite{Napoli} with an operational meaning in a phase discrimination task. Meanwhile, robustness is lower bounded by the relative entropy \cite{LamiPRA}. In the resource theory, the regularized relative entropy is equal to the smoothed version of the logarithmic robustness\cite{Gour}. 

We will work on an infinite-dimensional separable Hilbert space $\mathcal{H}$. We use $\mathcal{D(\mathcal{H})}$ for the set of density operators.
The robustness of a quantum state $\rho$ is defined as \cite{LamiPRA} \cite{LamiPRL} (It is added by 1 on the original definition \cite{Vidal})
    \begin{equation}\label{we1}
      R_{\mathcal{F}}(\rho):=\inf\{1+\lambda| \rho+\lambda\tau=(1+\lambda)\sigma, \tau\in\mathcal{D(\mathcal{H})}, \sigma\in \mathcal{F}\},
    \end{equation}
where $\mathcal{F}\subseteq \mathcal{D(\mathcal{H})}$ is the free state set. Resource theory is to split the state space into two parts: resource state set and free state set according to some specific properties of the states. Two reasonable assumptions about the $\mathcal{F}$ set are: that it is convex and that it is closed.

Usually, a free state is easily prepared in experiment in contract to a resource state. In quantum optics, coherent states are easily obtained in experiment and they resemble classical states in some sense so that it is reasonable to consider them as free states.
For continuous variable systems, the resource is called nonclassicality if the free state set is assumed to be $\mathcal{C}$, the set of (optical) coherent states and their probability mixture. Namely, a free state is defined as
\begin{equation}\label{we2}
 \sigma=\int P(\alpha)|\alpha\rangle\langle\alpha|[\frac{d^2\alpha}{2\pi}],
\end{equation}
 where $|\alpha\rangle$ is the coherent state and $P(\alpha)$ is a probability distribution function.  Nonclassicality is an idea with long history dated back to Glauber \cite{Glauber} and Sudarshan \cite{Sudarshan}. Any quantum state can be expressed in the form of (\ref{we2}), with $P(\alpha)$ now is called Glauber-Sudarshan P function, it is a quasi-probability instead of probability in (\ref{we2}). For a nonclassical state, the quasi-probability should have negative value at somewhere. However, due to the highly singularity of the quasi-probability, it is not easy to exhibit the negativity with experiment data at hand. Various methods have been developed to decide whether a quantum state is nonclassical, among them are the ``s-parametrized" quasi-probabilities including Wigner function which is widely used in experiment to elucidate nonclassicality. Experimental certification of noncalssicality with phase-space inequalities was also proposed\cite{Biagi}, recently. The quantification of noncalssicality \cite{Tan2020}\cite{Kwon} provides not only the quantitative aspect of the quantum resource of nonclassicality, but also helps to clarify the qualitative part of it. For examples, Fock states are classified as nonclassical because their quasi-probabilities are more singular than delta function. Now they are nonclassical due to positive values of nonclassicality.

   Gaussian states are important in continuous variable quantum information not only for experiment accessible but also for theoretical simplicity. Gaussian states have many applications in quantum information processing. We will mainly consider nonclassicality of quantum Guassian states in the following.



{\it Robustness of nonclassicality.} -
The definition of the robustness of resource has a dual space representation which leads to a lower bound:\cite{LamiPRA}
\begin{equation}\label{we3}
  R_{\mathcal{F}}(\rho)\geq\frac{\langle \rho,\omega \rangle}{\sup_{\sigma}\langle\sigma, \omega\rangle},
\end{equation}
for $\omega,\rho \in \mathcal{D(\mathcal{H})}$. We may call $\omega$ witness state since it witness resouce of the state $\rho$ when $\langle \rho,\omega \rangle > \sup_{\sigma}\langle\sigma, \omega\rangle$.

It has been known that robustness of resource for a pure state $|\psi\rangle$ is up bounded by $\langle \psi|\sigma^{-1}|\psi\rangle$, under the  restriction of $|\psi\rangle \in {\rm dom}({\sigma^{-\frac{1}{2}}})$, with dom denoting the domian, and $\sigma\in \mathcal{F}$ is a free state. With the up and lower bounds, the formulas for robustness of naclassicality have been given for Fock state, squeezed state and cat state when the up and lower bounds coincide with each other.

The definition of robustness leads to an upper bound of it. Notice that $\tau$ in (\ref{we1}) is a density operator, so we have $\lambda\tau\geq0$, hence $(1+\lambda)\sigma\geq\rho$. Assuming that $\sigma^{-1}$ exists, then we have $(1+\lambda)\mathbb{1}\geq \sigma^{-\frac{1}{2}}\rho\sigma^{-\frac{1}{2}}$ where $\mathbb{1}$ is the identity operator. We may diagonalize $\sigma^{-\frac{1}{2}}\rho\sigma^{-\frac{1}{2}}$ and find its largest eigenvalue which is denoted as $\Lambda$. For a given free state $\sigma$, the minimal $1+\lambda$ is equal to $\Lambda$. For the whole free state set $\mathcal{F}$, we obtain the robustness
\begin{equation}\label{we4}
  R_{\mathcal{F}}(\rho)=\inf_{\sigma\in\mathcal{ F}} \{\Lambda|\Lambda=\sup({\rm eig}(\sigma^{-\frac{1}{2}}\rho\sigma^{-\frac{1}{2}})) \},
\end{equation}
where eig stands for eigenvalue.
Apart from $\sigma\in \mathcal{F}$, some further requirements should be added on $\sigma$ such that $(1+\lambda)\sigma-\rho\geq 0$ is true for some $\lambda$.

One method of calculating the robustness of resource is to maximize the right hand of (\ref{we3}) with respect to all possible witness states. Another methods is using (\ref{we4}). Each of them is an optimization over quantum state set that could be quite involved since the set may be very large to contain all possible quantum states. A straightforward way is to guess a sufficiently small subset for witness states and another subset for free states, then calculate the lower and up bounds of the robustness. If the bounds coincide with each other, then we have the correct robustness.

{\it Nonclassicality of Gaussian states.} -
  For a multimode quantum state $\rho$,  its mean vector is $m=Tr(\rho \hat{R})$ and its covariance matrix $\gamma_{i,j}:=Tr[\rho(\hat{R}_{i}-m)(\hat{R}_{j}-m)]$, where $\hat{R}=(\hat{x}_{1},\hat{x}_{2},...,\hat{x}_{n},\hat{p}_{1},\hat{p}_{2},...,\hat{p}_{n})$. Where $\hat{x}_{j},\hat{p}_{j}$ are position and momentum operators of $j^{th}$ mode, respectively. A Gaussian state is completely characterized with its mean and covariance matrix.

The definition (\ref{we1}) of nonclassicality yields the following criterion of classicality.
\begin{proposition}\label{proposition1}
Let $\gamma$ be the CM of a general continuous variable classical state, then
\begin{equation}\label{we6}
  \gamma-I\geq0,
\end{equation}
Conversely, if this condition is satisfied, the Gaussian state with CM $\gamma$ is classical.
\end{proposition}
For any continuous variable quantum state, violation of this condition implies nonclassicality.

Proof: For the first statement, it had been shown that\cite{WernerWolf} if a state with CM $\gamma$ can be decomposed as convex combination of states with $\gamma_{k}$ and weight $\lambda_{k}$, then $\gamma-\Sigma_{k}\lambda_{k}\gamma_{k}\geq 0$. A classical state can be decomposed as the convex combination of coherent states as displayed in (\ref{we2}), the definition of classical state. Notice that the CM of any coherent state $|\alpha\rangle$ is $I$ regardless of $\alpha$. We denote the CM of coherent state as $\gamma_{\alpha}$, then $\int[\frac{d^2\alpha}{2\pi}]P(\alpha)\gamma_{\alpha}=\int[\frac{d^2\alpha}{2\pi}]P(\alpha)I=I$. So $\gamma-I$ is positive semi-definite.

For the converse statement, let $\gamma_{0}=\gamma-I$, then $\gamma_{0}$ is the covariance matrix of a classical Gaussian probability distribution P.
More explicitly, P function is the Fourier transformation of ${\rm Tr}[\rho D(\mu)]e^{\frac{|\mu|^2}{2}}$ \cite{KCTan2019} which is the characteristic function with covariance matrix $\gamma-I$ for a Gaussian state. Thus the P function is a positive Gaussian function\cite{Duan}. $\Box$

For a single mode squeezed state, we have $\gamma={\rm diag}(e^{2r},e^{-2r})$, the nonclassicality can be detected by the violation of criterion (\ref{we6}) even for infinitive small squeezing parameter. Notice that the nonclassicality of a squeezed state is not detected by Wigner function.

\begin{corollary}\label{corollary1}
   An orthogonal transformation $O$ in phase space does not change the classicality of a quantum Gaussian state.
\end{corollary}
   Proof: If a quantum Gaussian state with CM $\gamma$ is classical, then $\gamma-I\geq 0$ from Proposition \ref{proposition1}. Then $O\gamma O^T-I=O(\gamma -I)O^T\geq 0$. The equality comes from the orthogonality $OO^T=I$, the inequality comes from that orthogonal transformation preserves the positivity of a symmetric matrix. The transformed quantum Gaussian state with CM $O\gamma O^T$ is classical since $O\gamma O^T-I\geq0$.$\Box$

For a Gaussian state $\rho$, it is reasonable to assume the witness state $\omega$  and the free state $\sigma$ to be Gaussian. Without loss of generality, we may set the first moments of $\rho$ to be zeros. Then $\rho$ is characterized by its CM $\gamma$. Let $\omega$ be a zero mean Gaussian state with CM $\gamma_{\omega}$, then the robustness is lower bounded as
\begin{equation}\label{we7}
  R_{\mathcal{C}}(\rho)\geq \sqrt\frac{\det(\gamma_{\omega}+I)}{\det(\gamma_{\omega}+\gamma)}.
\end{equation}

The free state $\sigma$ can be chosen as lying on the border of classical state sets. Denote the CM of a Gaussian free state $\sigma$ to be $\gamma_{\sigma}$. Then we have $\gamma_{\sigma}-I\geq 0$, and
\begin{equation*}\label{we8}
\det(\gamma_{\sigma}-I)=0.
\end{equation*}
 This later condition shows us that $\sigma$ is on the border of free state set.
 Every CM can be diagonalized with proper symplectic transformation according to Williamson theorem \cite{Williamson}. That is $\gamma=\mathcal{S}\gamma_{d}\mathcal{S}^T$, where $\gamma_{d}={\rm diag}(\nu_{1}\nu_{2},...,\nu_{n},\nu_{1},\nu_{2},...,\nu_{n})$ with $\nu_{i}\geq 1$ (i=1,2,...,n) being the symplectic eigenvalues. The symplectic transformation $\mathcal{S}$ (explicitly worked out in \cite{Pirandola}) should fulfill $\mathcal{S}\Delta \mathcal{S}^T=\Delta$, where $\Delta=\left( \begin{array}{cc}
   0 & -I \\
   I & 0 \\
 \end{array}
\underline{}\right)
 $. The CM $\gamma_{d}$ corresponds a product thermal state $\rho_{th}=\bigotimes_{i=1}^{n}\tau_{\nu_{i}}$, with single mode thermal state $\tau_{\nu}=\frac{2}{\nu+1}\sum\emph{}_{k}(\frac{\nu-1}{\nu+1})^{k}|k\rangle\langle k|$.  The symplectic transformation $\mathcal{S}$ induces a unitary operation $U$. So we can write $\rho=U\rho_{th}U^{\dagger}$ and similarly $\sigma=U_{F}\sigma_{th}U_{F}^{\dagger}$. Denote $\rho'=U_{F}^{\dagger}\rho U_{F}$, then the eigenvalue of $\sigma^{-\frac{1}{2}}\rho\sigma^{-\frac{1}{2}}$ is equal to the eigenvalue of $\varrho =:\sigma_{th}^{-\frac{1}{2}}\rho'\sigma_{th}^{-\frac{1}{2}}$. Thus from (\ref{we4}), we have
 \begin{equation}\label{we8a}
 R_{\mathcal{F}}(\rho)\leq\Lambda=\sup[{\rm eig}(\varrho)].
\end{equation}
Using Fock space, we show in \cite{Supplementary} that $\varrho$ is an unnormalized Gaussian state when its eigenvalues are bounded.

Let the CM of $\rho'$ be $\gamma'$, the CCM be $\tilde{\gamma}'=L\gamma'L^{T}$, with $L=\frac{1}{\sqrt{2}}\left( \begin{array}{cc}
   -iI & -I \\
   iI & -I \\
 \end{array}
\underline{}\right)
 $. In Fock basis, the Gaussian state $\rho'$ is with matrix element \cite{Chen07}
 \begin{equation}\label{we9}
   \rho'_{l,m}=\mathcal{N}'\mathcal{O}_{l,m}(t,t')e^{\frac{1}{2}(t,t')(\sigma_{1}\otimes I+\beta')(t,t')^T},
 \end{equation}
 where $\mathcal{O}_{l,m}(t,t')=\frac{1}{\sqrt{\prod_{i=1}^{n}l_{i}!m_{i}!}}(\partial t)^{l}(\partial t')^{m}|_{t=t'=0}$ , Here $l=(l_{1},l_{2},...,l_{n}), m=(m_{1},m_{2},...,m_{n})$ and $(\partial t)^{l}$ stands for $\prod_{i=1}^{n}(\partial t_{i})^{l_{i}}$ and $(\partial u)^v=\frac{\partial^{v}}{\partial u^v} $. Here $\beta'=(\sigma_{3}\otimes I)(\frac{\tilde{\gamma}}{2}'+\sigma_{1}\otimes \frac{I}{2})^{-1}(\sigma_{3}\otimes I)$,
 $\sigma_{i} (i=1,2,3)$ are Pauli matrices. The normalization is $\mathcal{N}'=|\det\beta'|^{\frac{1}{2}}=2^{n}\det(\gamma'+I)^{-\frac{1}{2}}$.
The product thermal state $\sigma_{th}=\bigotimes_{i}^{n}(1-u_{i}) \sum_{k_{i}=0}^{\infty} u_{i}^{k_{i}}|k_{i}\rangle\langle k_{i}|$ is specified by $u_{i}=\frac{\nu_{i}^{\sigma}-1}{\nu_{i}^{\sigma}+1}$, where $\nu_{i}^{\sigma}$ are the symplectic eigenvalues of free Gaussian state $\sigma$.

The matrix element of $\varrho$ is
\begin{equation}\label{we10}
  \varrho_{l,m}=\prod_{i,j=1}^{n}\frac{1}{\sqrt{(1-u_{i})(1-u_{j})}}u_{i}^{-\frac{l_{i}}{2}}\rho'_{l,m}u_{j}^{-\frac{m_{j}}{2}}.
\end{equation}
Our aim is to transform $\varrho_{l,m}$ into the form of (\ref{we9}), although with different parameters.
Let $w=(w_{1},w_{2},...,w_{n}), w'=(w'_{1},w'_{2},...,w'_{n})$,  with $w_{i}=\sqrt{u_{i}}t_{i},w'_{i}=\sqrt{u_{i}}t'_{i} $, then
\begin{eqnarray}\label{we11}
&&  \varrho_{l,m}=\frac{\mathcal{N}'}{\prod_{i}(1-u_{i}) }\mathcal{O}_{l,m}(w,w')e^{\frac{1}{2}(t,t')(\sigma_{1}\otimes I+\beta')(t,t')^T} \nonumber\\
&&  =\frac{\mathcal{N}'}{\prod_{i}(1-u_{i}) }\mathcal{O}_{l,m}(w,w')e^{\frac{1}{2}(w,w')\Gamma(\sigma_{1}\otimes I+\beta')\Gamma(w,w')^T} \nonumber\\
&&=\frac{1}{\prod_{i}(1-u_{i})}\frac{\mathcal{N}'}{\mathcal{N}''}\rho''
\end{eqnarray}
 where $\Gamma={\rm diag}(\frac{1}{\sqrt{u_{1}}},\frac{1}{\sqrt{u_{2}}},...,\frac{1}{\sqrt{u_{n}}},\frac{1}{\sqrt{u_{1}}},\frac{1}{\sqrt{u_{2}}},...,\frac{1}{\sqrt{u_{n}}})$ and.
It is clear that $\rho''$ is a Gaussian state with CCM $\tilde{\gamma}''$ which is defined by $\beta''=(\sigma_{3}\otimes I)(\frac{\tilde{\gamma}''}{2}+\sigma_{1}\otimes \frac{I}{2})^{-1}(\sigma_{3}\otimes I)$ and $\beta''+\sigma_{1}\otimes I=\Gamma(\beta'+\sigma_{1}\otimes I)\Gamma$. The normalization $\mathcal{N}''=|\det\beta''|^{\frac{1}{2}}=2^{n}\det(\gamma''+I)^{-\frac{1}{2}}$. The symplectic eigenvalues $\nu_{i}''$ of $\rho''$ can be found from its CM $\gamma''$. The eigenvalues of $\Delta^{-1}\gamma''$ are $\pm i \nu_{i}''$ (i=1,...,n). The Gaussian state $\rho''$ can be unitarily transformed to a multimode product thermal state $\rho''_{th}$ which has a largest eigenvalue $\prod_{i=1}^n\frac{2}{\nu''_{i}+1}$.  Clearly the largest eigenvalue of $\rho''$ is just the largest eigenvalue of the product thermal state $\rho''_{th}$. Hence the largest eigenvalue of $\varrho$ is
\begin{equation}\label{we13}
  \Lambda=\prod_{i}\frac{2}{(1-u_{i})(\nu''_{i}+1)}\sqrt{\frac{\det(\gamma''+I)}{\det(\gamma'+I)}}.
\end{equation}
It is an upper bound for the robustness of nonclassicality. Minimizing $\Lambda$ with respect to free state set will lead to a tight upper bound. Keep in mind that the eigenvalues of $\rho''$ should be bounded such that $Tr\rho''=1$. This limits the range of free state $\sigma$. Notice that $\sigma=U_{F}\sigma_{th}U_{F}^{\dagger}$, the thermal state $\sigma_{th}$ and the unitary transform $U_{F}$ are related since $\sigma$ is a free state. This restriction on $\sigma_{th}$ and $U_{F}$ should be considered when $\Lambda$ is minimized. The parameters of $\sigma_{th}$ appear in $u_{i}, \gamma''$, while the parameters of $U_{F}$ appear in $\gamma',\gamma''$.

For the upper and lower bounds, we have the following corollary.
\begin{corollary}\label{corollary2}
An orthogonal transform in phase space does not change the lower and up bounds of the robustness of nonclassicality for a Gaussian state.
\end{corollary}
Proof: Let the associated unitary transformation be $U$, since $U\sigma U^{\dagger}$ is classical for a classical state $\sigma$ according to Corollary \ref{corollary1}, that the lower bound for robustness of nonclassicality is not changed by the unitary transform can be easily seen from the definition of lower bound. The upper bound does not change since $(U\sigma U^{\dagger})^{-\frac{1}{2}}$=$U\sigma^{-\frac{1}{2}}U^{\dagger}$. $\Box$


{\it Entanglement of Gaussian states.} -
There are many entanglement criteria for Gaussian states\cite{Duan}\cite{Simon}\cite{WernerWolf}\cite{Chen07}\cite{Chen23PRA}\cite{ChenPRR}, in contrast to the calculable entanglement measures which are seldom proposed for Gaussian states. The entanglement of formation is one of the tractable examples for bipartite ($1\times1$) symmetric Gaussian states\cite{Giedke}\cite{Marian}\cite{Ralph}\cite{Lami2019}, also there are some results on entanglement of formation for high-dimentional entanglement\cite{Gisin} and multipartitie Gaussian states\cite{Adesso2010}\cite{Adesso2007}. The robustness of entanglement for a Gaussian state will be illustrated in the following. We will study the lower and upper bounds of the robustness of entanglement. The formulas (\ref{we3})(\ref{we4}) are also true for robustness of entanglement, with the free state set $\mathcal{F}$ being the separable state set instead of coherent state set in the robustness of nonclassicality. The change of free state set has a profound effect. In evaluating lower bound of robustness of entanglement with (\ref{we3}), it is reasonable to assume the witness state $\omega$ to be Gaussian for a Gaussian state $\rho$. The problem is to evaluate the supreme mean of $\omega$ over product state set. Our conjecture is that the supreme mean over product state set is achieved by product Gaussian states (similar Gaussian extremality had been studied in Ref. \cite{Wolf}). It is numerically proven to be true for a $1\times1$ witness state $\omega$ with standard form of CM \cite{Chen23PRA}. Using Bosonic Gaussian channel, we also show the conjecture for $1\times1$ system to be true with the assumption of $\omega\rightarrow \infty $ ( all the parameters of its CM $\rightarrow \infty$)\cite{ChenPRR}. 
Then it follows from (\ref{we3}) the lower bound of robustness of entanglement
\begin{equation}\label{we13a}
   R_{\mathcal{E}}(\rho)\geq \sqrt\frac{\inf_{\gamma_{ps}}\det(\gamma_{\omega}+\gamma_{ps})}{\det(\gamma_{\omega}+\gamma)}.    
\end{equation}
with $\gamma_{ps}={\rm diag}(x,y,1/x,1/y)$ the CM of product squeezed state.
Also, for a three mode symmetric state, we have shown numerically that its supremal mean over product of three single-mode state is achieved by three fold product of a squeezed state \cite{ChenPRR}. Thus the lower bound of robusteness of entanlgment  can be written with CMs just as in (\ref{we13a}), with $\gamma_{ps}={\rm diag}(x,x,x,1/x,1/x,1/x)$.

{\it Examples and Applications.} -
   The robustness of nonclassicality can be explicitly obtained as an illustration of the method in the former text with the examples of single mode and two-mode Gaussian states and some multi-mode symmetric Gaussian states.
   First we consider a squeezed thermal state $\rho=S(r)\tau_{\nu}S^{\dagger}(r)$, its robustness of nonclassicality is \cite{Supplementary}
 \begin{equation}\label{we11a}
   R_{\mathcal{C}}(\rho)=\max\{\frac{e^{r}}{\sqrt{\nu}},1\}.
 \end{equation}
 where $r$ is the squeezing parameter of squeezing operator $S(r)$ and $N$ is the average photon number of the thermal noise state $\tau_{\nu}$. If we define $\eta=\frac{1}{2}\log(\nu)$, then the logrithmic robustness is
  $\log(R_{\mathcal{C}})=r-\eta$ when $r\geq \eta$. It is the difference of squeezing parameter and noise parameter. The nonclassicality of the state increases with squeezing and decreases with noise.

Any two-mode Gaussian state can be transformed to a standard form Gaussian state by local unitary operations. For simplicity, we will consider the standard form of two-mode Gaussian states in the following. The CM of a two-mode standard form Gaussian state is
\begin{equation}\label{we21}
  \gamma=\gamma_{x}\oplus \gamma_{p},
\end{equation}
with $\gamma_{x}=\left(
                   \begin{array}{cc}
                     a & c_{1} \\
                     c_{1} & b \\
                   \end{array}
                 \right)$ and $\gamma_{p}=\left(
                                            \begin{array}{cc}
                                              a & -c_{2} \\
                                              -c_{2} & b \\
                                            \end{array}
                                          \right)$.
For a two-mode Gaussiam state with the standard form of CM, the upper bound of nonclassicality can be directly calculated with (\ref{we13}) \cite{Supplementary}. For the lower bound, we develop two lower bounds from (\ref{we7}) \cite{Supplementary}, the lower bound 1 in Fig.1 (a) is analytical, the lower bound 2 is numerical. We draw lower bound 2 when lower bound 1 does not fit the upper bound. The lower bound 2 always fits with the upper bound well. 

Similarly, the upper bound of entanglement can be derived from (\ref{we13})\cite{Supplementary}. We also have two kind of lower bounds from (\ref{we13a}), lower bound 1 in Fig.1(b) is analytical, lower bound 2 is numerical. The upper and lower bounds fit with each other well.
 
A special case of standard two-mode Gaussian state is the two-mode squeezed thermal state with $c_{1}=c_{2}=c$ in (\ref{we21}) for its CM. The state can be written as $\rho=S_{2}(r)\tau_{\nu_{1}}\otimes\tau_{\nu_{2}}S_{2}^{\dagger}(r)$, where the two-mode squeeze operator $S_{2}(r)=\exp[r(a^{\dagger}_{1}a^{\dagger}_{2}-a_{1}a_{2})]$ and $a_{i},a^{\dagger}_{i}$ (i=1,2) are the annihilation and creation operators,respectively. The upper and lower bounds of robustness of nonclassicality coincide with each other. The robustness of nonclassicality is \cite{Supplementary}
\begin{equation}\label{we21a}
 R_{\mathcal{C}}(\rho)=\left \{\begin{array}{l}
                              \frac{2e^{2r}}{\nu_{1}+\nu_{2}}, \text{ \quad  \quad  \quad  for } a-c<1, b-c<1,\\
                              1 ,   \text{\quad \quad \quad \quad  \quad for } \delta=(a-1)(b-1)-c^2\geq0,\\
                              \frac{2(a-1)}{(a-1)(b+1)-c^2}, \text{ for } a-c\geq1, b-c<1, \delta<0,\\
                              \frac{2(b-1)}{(a+1)(b-1)-c^2}, \text{ for } a-c<1, b-c\geq1, \delta<0.
                            \end{array}\right.
\end{equation}
The robustness of entanglement of the two-mode squeezed thermal state is equal to its robustness of nonclassicality, $R_{\mathcal{E}}(\rho)=R_{\mathcal{C}}(\rho)$, this is due to the fact that in obtaining $R_{\mathcal{E}}(\rho)$ we do not need the local squeezings (the witness $\omega$ is also a two-mode squeezed thermal state, thus $x=y=1$ in (\ref{we13a})).

\begin{figure}
\centering
\subfigure[\label{Fig.1a}]{
\includegraphics[width=1.65in]{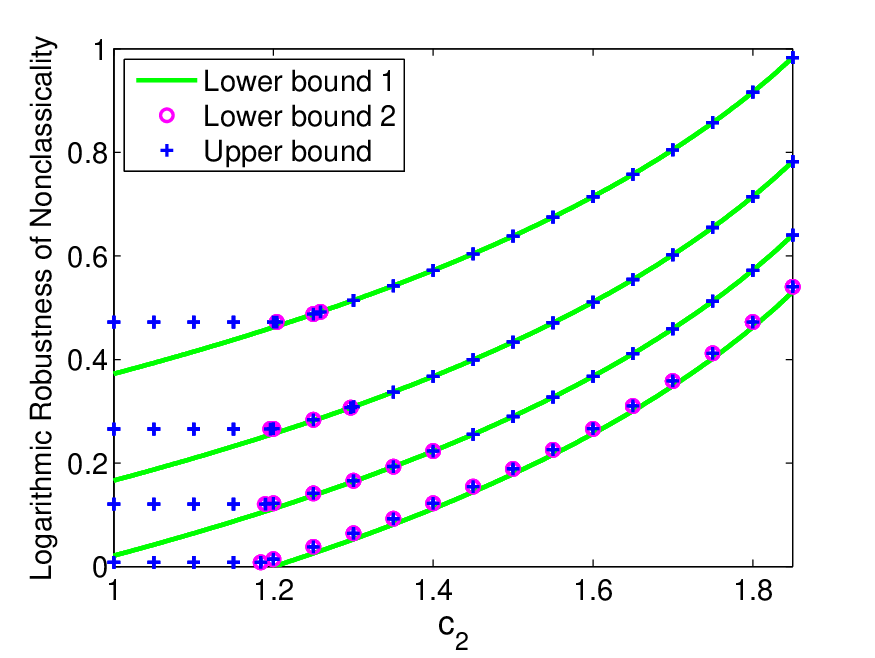}}
\subfigure[\label{Fig.1b}]{
\includegraphics[width=1.65in]{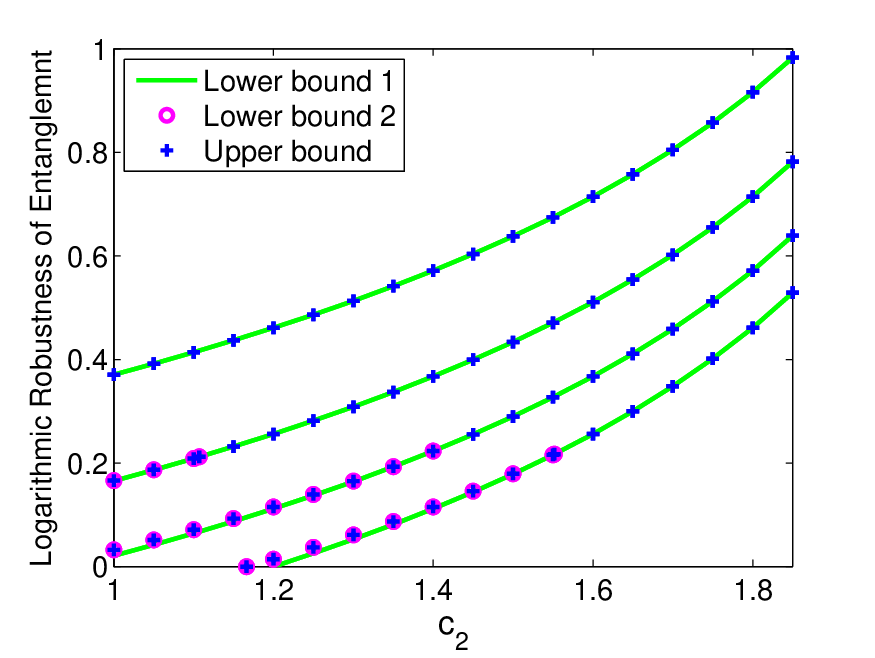}}
\caption{(a) Logarithmic robustness of nonclassicality for two-mode standard form Gaussian states with parameters $a=2.4, b=2 $, and four cases of $c_{1}$. From left top to right bottom, $c_{1}=1.8,1.6,1.4,1.2$, respectively.
(b)Logarithmic robustness of entanglement for two-mode standard form Gaussian states with parameters $a=2.4, b=2 $, and four cases of $c_{1}$. From left top to right bottom, $c_{1}=1.8,1.6,1.4,1.2$, respectively.}
\end{figure}
\label{Fig1}


Let's consider the robustness of nonclassicality for an $n$-mode symmetric Gaussian state with the CM of form $\gamma=\gamma_{x}\oplus\gamma_{p}$, with the diagonal and off-diagonal elements of $\gamma_{x}$ ($\gamma_{p}$) are $a, c_{1}$ ($b, -c_{2}$), respectively.

Matrix $\gamma_{x}$ commutates with matrix $\gamma_{p}$. Thus they have the same orthonormal basis. The pile of these row bases is an orthogonal $n\times n$ matrix $X$ such that $X\gamma_{x}X^{T}$ and $X\gamma_{p}X^{T}$ are  diagonal.

With the orthogonal $X$, we have a symplectic transform $\mathcal{S}=X\oplus X$ due to the fact that $\mathcal{S}\Delta \mathcal{S}^T=\Delta$. Thus $\mathcal{S}$ is an orthogonal and symplectic tranform. The upper and lower bounds of robustness of nonclassicality will keep invariant under such a symplectic transform according to \ref{corollary2}. The problem of the bounds for an $n$-mod symmetric Gaussian state is transformed to that of the bounds for a product Gaussian state with CM of $\mathcal{S}\gamma \mathcal{S}^{T}={\rm diag}[a+(n-1)c_{1},a-c_{1},...,a-c_{1},b-(n-1)c_{2},b+c_{2},...,b+c_{2}]$. The problem is reduced to that of single-mode Gaussian state. for an $n$-mod symmetric Gaussian state $\rho$, we have the robustness of nonclassicality
\begin{eqnarray}\label{we21c}
  R_{\mathcal{C}}(\rho)=\max\{\frac{1}{\sqrt{a+(n-1)c_{1}}},\frac{1}{\sqrt{b-(n-1)c_{2}}},1\}\nonumber\\
  \times\max\{\frac{1}{\sqrt{a-c_{1}}^{n-1}},\frac{1}{\sqrt{b+c_{2}}^{n-1}},1\}
\end{eqnarray}

For continuous variable (CV) GHZ state, $a=\frac{1}{n}[e^{2r}+(n-1)e^{-2r}]$, $c_{1}=c_{2}=\frac{1}{n}[e^{2r}-e^{-2r}]$, $b=\frac{1}{n}[e^{-2r}+(n-1)e^{2r}]$,the robustness of nonclassicality is
\begin{equation}\label{we21d}
  R_{\mathcal{C}}(\rho_{GHZ})=e^{nr}.
\end{equation}

The robustness of entanglement for the  CV-GHZ  state is very interesting since it is equal to $e^{2r}$ for either two-mode CV-GHZ state or three-mode CV-GHZ state. All the n-mode CV-GHZ states has the same robustness of entanglement  upper bound $e^{2r}$ \cite{Supplementary}.
\begin{equation}\label{we21e}
  R_{\mathcal{E}}(\rho_{GHZ})\leq e^{2r},   {\text  \quad {\rm regardless \quad of}\quad n}.
\end{equation}
The fully separable state $\sigma_{GHZ}$ for calculating the upper bound is with a CM of $\gamma_{\sigma_{GHZ}}=\mathcal{S}^{-1}[{\rm diag}(e^{4s_{1}-2r},e^{2r})\otimes{\rm diag}(e^{-2r},e^{2r})^{\otimes n-1}]\mathcal{S}^{-T}$, where $\mathcal{S}$ is the symplectic transform that diagonalizes $\gamma_{\sigma_{GHZ}}$ and $s_{1}=\frac{1}{4}\log(2e^{4r}-1)$.
If the conjecture of superemal Gaussian mean over product single-mode states being achieved by product Gaussian state is true for all n-mode symmetric system, then the lower bound $e^{2r}$ can be derived, we then have the exact robustness of entanglement for CV-GHZ state.
At present, the conjecture has been numerically verified for three-mode symmetric Gaussian states\cite{ChenPRR}.  

 Entanglement is significantly different from nonclassicality for multipartite sytstem. The reason behind is that entanglement is free of local squeezings, these local squeezings reduce the entanglement resources\cite{Supplementary}.

{\it Conclusion.}- 
The resources in a Gaussian quantum state can be quantified with its lower and upper bounds of robustness when the bounds are equal. We turn the definition of nonclassicality for a quantum state into its covariance matrix form, and prove that the nonclassicality of a Gaussian state does not changed by orthogonal symplectic transform, one of its direct application is to decompose the nonclassicality of a symmetric Gaussian state to that of product single-mode squeezed states. We give the robusteness of nonclassicality for single-mode, two-mode standard form Gaussian states and multimode symmetric Gaussian states. The robustness of entanglement is also obtained for two-mode standard form Gaussian states and multimode CV-GHZ states. Although in two-mode case, the nonclassicality only a little bit larger than entanglement measured with robustness (they are equal for a squeezed thermal state), they are significantly different for a multimode CV-GHZ state. The logarithmic robustness of nonclassicality increases with the number of modes, the logarithmic robustness of entanglement keeps constant regardless of the number of modes, it is reminiscent of the monogamy of entanglement. The important contribution of this letter is to propose the upper bound of robustness for mixed Gaussian states. The upper bound proposed here should not be limited to Gaussian states, it is quite universal.

{\it Acknowledgment.} -
This work is supported by the National Natural Science Foundation  of China (Grant No.61871347)


\renewcommand\refname{------------------------------}
\bibliographystyle{unsrt}
\bibliography{RNCGbibfile}

\begin{thebibliography}{10}

\bibitem{Bennett}
C.~H. Bennett, D.~P. DiVincenzo, J.~A. Smolin, and W.~K. Wootters.
\newblock Mixed-state entanglement and quantum error correction.
\newblock {\em Phys. Rev. A}, 54:3824, 1996.

\bibitem{Vedral}
V.~Vedral, M.~B. Plenio, M.~A. Rippin, and P.~L. Knight.
\newblock Quantifying entanglement.
\newblock {\em Phys. Rev. Lett.}, 78:2275, 1997.

\bibitem{Takagi}
R.~Takagi, B.~Regula, K.~F. Bu, Z.-W. Liu, and G.~Adesso.
\newblock Operational advantage of quantum resources in subchannel
  discrimination.
\newblock {\em Phys. Rev. Lett.}, 122:140402, 2019.

\bibitem{Piani}
M.~Piani and J.~Watrous.
\newblock Necessary and sufficient quantum information characterization of
  einstein-podolsky-rosen steering.
\newblock {\em Phys. Rev. Lett.}, 114:060404, 2015.

\bibitem{LamiPRA}
L.~Lami, B.~Regula, R.~Takagi, and G.~Ferrari.
\newblock Taming the infinite: framework for resource quantification in
  infinite-dimensional general probabilistic theories.
\newblock {\em Phys. Rev. A}, 103:032424, 2021.

\bibitem{Vidal}
G.~Vidal and R.~Tarrach.
\newblock Robustness of entanglement.
\newblock {\em Phys. Rev. A}, 59:141--155, 1999.

\bibitem{Steiner}
M~Steiner.
\newblock Generalized robustness of entanglement.
\newblock {\em Phys. Rev. A}, 67:054305, 2003.

\bibitem{Napoli}
C.~Napoli, T.~R. Bromley, M.~Cianciaruso, M.~Piani, N.~Johnston, and G.~Adesso.
\newblock Robustness of coherence: An operational and observable measure of
  quantum coherence.
\newblock {\em Phys. Rev. Lett.}, 122:140402, 2019.

\bibitem{Gour}
F.~Brandão and G.~Gour.
\newblock Reversible framework for quantum resource theories.
\newblock {\em Phys. Rev. Lett.}, 115:070503, 2015.

\bibitem{LamiPRL}
B.~Regula, L.~Lami, G.~Ferrari, and R.~Takagi.
\newblock Operational quantification of continuous-variable quantum resources.
\newblock {\em Phys. Rev. Lett.}, 126:110403, 2021.

\bibitem{Glauber}
R.~J. Glauber.
\newblock The quantum theory of optical coherence.
\newblock {\em Phys. Rev}, 130:2529--2539, 1963.

\bibitem{Sudarshan}
E.~C.~G. Sudarshan.
\newblock Equivalence of semiclassical and quantum mechanical descriptions of
  statistical light beams.
\newblock {\em Phys. Rev. Lett}, 10:277--279, 1963.

\bibitem{Biagi}
N.~Biagi, M.~Bohmann, E.~Agudelo, M.~Bellini, and A.~Zavatta.
\newblock Experimental certification of nonclassicality via phase-space
  inequalities.
\newblock {\em Phys. Rev. Lett.}, 126:023605, 2021.

\bibitem{Tan2020}
K.~C. Tan, S.~Choi, and Jeong H.
\newblock Negativity of quasiprobability distributions as a measure of
  nonclassicality.
\newblock {\em Phys. Rev. Lett.}, 124:110404, 2020.

\bibitem{Kwon}
H.~Kwon, K.~C. Tan, T.~Volkoff, and H~Jeong.
\newblock Nonclassicality as a quantifiable resource for quantum metrology.
\newblock {\em Phys. Rev. Lett.}, 122:040503, 2019.

\bibitem{WernerWolf}
R.~F. Werner and M.~M. Wolf.
\newblock Bound entangled gaussian states.
\newblock {\em Phys. Rev. Lett}, 86:3658, 2001.

\bibitem{KCTan2019}
K.~C. Tan and H.~Jeong.
\newblock Nonclassical light and metrological power: An introductory review.
\newblock {\em {AVS} Quantum Science}, 1:014701, 2019.

\bibitem{Duan}
Lu-Ming Duan, G.~Giedke, J.~I. Cirac, and P.~Zoller.
\newblock Inseparability criterion for continuous variable systems.
\newblock {\em Phys. Rev. Lett}, 84:2722, 2000.

\bibitem{Williamson}
J.~Williamson.
\newblock On the algebraic problem concerning the normal forms of linear
  dynamical systems.
\newblock {\em Am. J. Math.}, 58:141--163, 1936.

\bibitem{Pirandola}
J.~L. Pereira, L.~Banchi, and S.~Pirandola.
\newblock Symplectic decomposition from submatrix determinants.
\newblock {\em Proc. R. Soc. A.}, 477:20210513, 2021.

\bibitem{Supplementary}
See supplemental material at [url will be inserted by publisher].

\bibitem{Chen07}
X.~Y. Chen.
\newblock Fock-space inseparability criteria of bipartite continuous-variable
  quantum states.
\newblock {\em Phys. Rev. A}, 76:022309, 2007.

\bibitem{Simon}
R.~Simon.
\newblock Peres-horodecki separability criterion for continuous variable
  systems.
\newblock {\em Phys. Rev. Lett}, 84:2726, 2000.

\bibitem{Chen23PRA}
X.~Y. Chen, M.~Miao, R.~Yin, and J~Yuan.
\newblock Gaussian entanglement witness and refined werner-wolf criterion for
  continuous variables.
\newblock {\em Phys. Rev. A}, 107:022410, 2023.

\bibitem{ChenPRR}
X.~Y. Chen, M.~Miao, R.~Yin, and J~Yuan.
\newblock Bosonic gaussian channel and gaussian witness entanglement criterion
  of continuous variables.
\newblock {\em Phys. Rev. Research}, 5:033066, 2023.

\bibitem{Giedke}
G.~Giedke, M.~M. Wolf, O.~Krüger, R.~F. Werner, and J.~I. Cirac.
\newblock Entanglement of formation for symmetric gaussian states.
\newblock {\em Phys. Rev. Lett.}, 91:107901, 2003.

\bibitem{Marian}
P~Marian and T.~A. Marian.
\newblock Entanglement of formation for an arbitrary two-mode gaussian state.
\newblock {\em Phys. Rev. Lett.}, 101:220403, 2008.

\bibitem{Ralph}
Tserkis S., S.~Onoe, and T.~C. Ralph.
\newblock Quantifying entanglement of formation for two-mode gaussian states:
  Analytical expressions for upper and lower bounds and numerical estimation of
  its exact value.
\newblock {\em Phys. Rev. A}, 99:052337, 2019.

\bibitem{Lami2019}
L.~Lami, S.~Khatri, G.~Adesso, and M.~M. Wilde.
\newblock Extendibility of bosonic gaussian states.
\newblock {\em Phys. Rev. Lett.}, 123:050501, 2019.

\bibitem{Gisin}
A.~Martin, T.~Guerreiro, A.~Tiranov, S.~Designolle, F.~Fröwis, B.~Brunner,
  M.~Huber, and N.~Gisin.
\newblock Quantifying photonic high-dimensional entanglement.
\newblock {\em Phys. Rev. Lett.}, 118:110501, 2017.

\bibitem{Adesso2010}
G.~Adesso and A.~Datta.
\newblock Quantum versus classical correlations in gaussian states.
\newblock {\em Phys. Rev. Lett.}, 105:030501, 2010.

\bibitem{Adesso2007}
G.~Adesso and F.~Illuminati.
\newblock Strong monogamy of bipartite and genuine multipartite entanglement:
  The gaussian case.
\newblock {\em Phys. Rev. Lett.}, 99:150501, 2007.

\bibitem{Wolf}
M.~M. Wolf, G.~Giedke, and J.~I. Cirac.
\newblock Extremality of gaussian quantum states.
\newblock {\em Phys. Rev. Lett}, 96:080502, 2006.

\end{thebibliography}

\newpage
\setcounter{equation}{0}

\renewcommand\theequation{S.\arabic{equation}} 

\section*{Supplemental Material: Quantifying nonclassicality and entanglement of Gaussian states}

\subsection*{A. Single-mode Gaussian state}
A single mode Gaussian state $\rho$ is characterized by CM of
    \begin{equation}\label{we14}
      \gamma=\left(
               \begin{array}{cc}
                 a & c \\
                 c & b \\
               \end{array}
             \right)
    \end{equation}
  with the Schr\"{o}dinger-Robertson uncertainty relation of $\gamma+i\Delta \geq0$ which leads to $ab-c^2\geq 1$. The symplectic eigenvalue is $\nu=\sqrt{ab-c^2}$. The CM $\gamma$ can be written as diagonal CM $\nu I_{2}$ applied on with squeezing and rotation, where $I_{2}$ is the $2\times2$ identity matrix.

   In evaluating the lower bound of the robustness with (\ref{we7}), without loss of generality, we may omit the rotation according to Corollary \ref{corollary2} and assume the state to be $\rho=S(r)\rho_{th}(\nu)S^{\dagger}(r)$ where $\rho_{th}(\nu)$ is a single mode thermal state with symplectic eigenvalue $\nu$ or mean photon number $\frac{\nu-1}{2}$, the squeezing operator $S(r)=\exp[\frac{r}{2}(\hat{a}^{\dagger 2}-\hat{a}^{2})]$. The CM of state $\rho$ is $\gamma={\rm diag}(\nu e^{2r},\nu e^{-2r})$. Let the witness state be with CM $\gamma_{\omega}={\rm diag}(\nu_{\omega} e^{2r_{\omega}},\nu_{\omega} e^{-2r_{\omega}})$. Then
  \begin{equation}\label{15}
  \sup_{\gamma_{\omega}}\frac{\det(\gamma_{\omega}+I)}{\det(\gamma_{\omega}+\gamma)}=\lim_{r_{\omega}\rightarrow\infty}\frac{\cosh(2r_{\omega})}{\nu\cosh(2r_{\omega}-2r)}=\frac{e^{2r}}{\nu}.
  \end{equation}
 Hence we have
 \begin{equation}\label{we16}
   R_{\mathcal{C}}(\rho)\geq \frac{e^{r}}{\sqrt{\nu}}.
 \end{equation}

 The criterion (\ref{we6}) of classicality leads to $\nu e^{-2r}\geq 1$ for a single mode Gaussian state with CM $\gamma={\rm diag}(\nu e^{2r},\nu e^{-2r})$. A single mode Gaussian state on the boundary of classical state set is $\sigma_{s}$ with CM $\gamma={\rm diag}(\nu(s) e^{2s},\nu(s) e^{-2s})$ and $\nu(s)=e^{2s}$. Thus we have $\sigma_{s}=S(s)\tau_{\nu(s)}S^{\dagger}(s)$. Let $v_{s}=\frac{\nu(s)-1}{\nu(s)+1}=\tanh(s)$, the thermal state will be $\tau_{\nu(s)}=(1-v_{s})\sum_{k}v_{s}^{k}|k\rangle\langle k|$. For the up bound of robustness of state $\rho$, we consider the largest eigenvalue of $\sigma_{s}^{-\frac{1}{2}}\rho\sigma_{s}^{-\frac{1}{2}}$ which is equivalent to the largest eigenvalue of $\tau_{\nu(s)}^{-\frac{1}{2}}\rho'\tau_{\nu(s)}^{-\frac{1}{2}}$, where $\rho'=S^{\dagger}(s)\rho S(s)$. If the CM of single mode Gaussian state $\rho$ is $\gamma={\rm diag}(\nu e^{2r},\nu e^{-2r})$, then the CM of $\rho'$ is $\gamma'={\rm diag}(\nu e^{2(r-s)},\nu e^{-2(r-s)})$. We may express the Gaussian state $\rho'$ in Fock space with element \cite{Chen07}
\begin{eqnarray}\label{we17}
  \rho'_{l,m}=\sum_{k,i,j|k+2i=m,k+2j=l}\frac{\sqrt{(k+2i)!(k+2j)!}}{k!i!j!}\nonumber\\
\times \sqrt{A^2-B^2}(1-A)^{k}(\frac{B}{2})^{i+j}.
\end{eqnarray}
Where $   \left(
               \begin{array}{cc}
                 B & A \\
                 A& B\\
               \end{array}
             \right)=(\frac{\tilde{\gamma}'_{+}}{2})^{-1}$, with $\tilde{\gamma}'_{+}=L{\gamma}'_{+}L^{T}$ and ${\gamma}'_{+}=\gamma'+I$.
The unnormalized Gaussian state $\varrho$ has element
\begin{eqnarray}\label{we17a}
  \varrho_{l,m}=(1-v_{s})^{-\frac{1}{2}}v_{s}^{-\frac{l}{2}}\rho'_{l,m}v_{s}^{-\frac{m}{2}} (1-v_{s})^{-\frac{1}{2}}\nonumber\\
=\sum_{k,i,j|k+2i=m,k+2j=l}\frac{\sqrt{(k+2i)!(k+2j)!}}{k!i!j!}\nonumber\\
\times \frac{\sqrt{A^2-B^2}}{1-v_{s}}(\frac{1-A}{v_{s}})^{k}(\frac{B}{2v_{s}})^{i+j}.
\end{eqnarray}
Comparing (\ref{we17a}) with (\ref{we17}), we have $ \varrho=\frac{1}{1-v_{s}}\sqrt{\frac{A^2-B^2}{A'^2-B'^2}}\rho''$ as far as the Gaussian state $\rho''$ exists. Where $1-A'=\frac{1-A}{v_{s}}, B'=\frac{B}{v_{s}}$. The CM of $\rho''$ is $\gamma''=\gamma''_{+}-I$, $\gamma''_{+}=L^{\dagger}\tilde{\gamma}''_{+}L^{*}$ and  $ \tilde{\gamma}''_{+}=2\left(
               \begin{array}{cc}
                 B' & A' \\
                 A'& B'\\
               \end{array}
             \right)^{-1}$.
 Let $\nu''$ be the symplectic eigenvalue of the state $\rho''$, then $\nu''=\sqrt{1+\frac{4(1-A')}{A'^2-B'^2}}$. The largest eigenvalue of $\rho''$ is $\frac{2}{\nu''+1}$, it follows the largest eigenvalue of $\varrho$
\begin{eqnarray}\label{we17b}
  \Lambda=\frac{2}{(1-v_{s})(1+\nu'')}\sqrt{\frac{A^2-B^2}{A'^2-B'^2}}\nonumber\\
=\frac{2v_{s}\sqrt{A^2-B^2}}{1-v_{s}}[\sqrt{(v_{s}-1+A)^2-B^2}\nonumber\\
  +\sqrt{(v_{s}+1-A)^2-B^2}]^{-1},
\end{eqnarray}
Denote $\eta=\log\sqrt{\nu}$, then $A+B=\frac{2}{1+e^{2(\eta+s-r)}}, A-B=\frac{2}{1+e^{2(\eta+r-s)}}$. 
The largest eigenvalue of $\varrho$ can be written as
\begin{eqnarray}\label{we18}
  \Lambda=\frac{2\sinh(s)e^{-\eta}}{1-\tanh(s)}[\sqrt{\sinh(2s-\eta-r)\sinh(r-\eta)}\nonumber\\
  +\sqrt{\sinh(2s+\eta-r)\sinh(r+\eta)}]^{-1},
\end{eqnarray}
 After a lengthy but straightforward algebra, we have
\begin{equation}\label{we19}
  R_{\mathcal{C}}(\rho)\leq \inf_{s}(\Lambda)=e^{r-\eta}=\frac{e^{r}}{\sqrt{\nu}},
\end{equation}
when $r\geq \eta$. The lower bound from (\ref{we16}) and the up bound from (\ref{we19}) coincide with each other. So when $r\geq \log\sqrt{\nu}$ (namely $\gamma\leq I$), the robustness of nonclassicality is
\begin{equation}\label{we20}
  R_{\mathcal{C}}(\rho)=\frac{e^{r}}{\sqrt{\nu}}=\sqrt{\frac{2}{a+b-\sqrt{(a-b)^2+4c^2}}}.
\end{equation}

\subsection*{B. Two-mode Gaussian state}
For the lower bound of robustness, we consider witness state $\omega$ to be also a standard form of Gaussian state with CM, $\gamma_{\omega}=\gamma_{\omega x}\oplus\gamma_{\omega p}$, where $\gamma_{\omega x}=\left(
                                                               \begin{array}{cc}
                                                                 a_{\omega 1} & c_{\omega 1} \\
                                                                 c_{\omega 1} & b_{\omega 1} \\
                                                               \end{array}
                                                             \right)$ and $\gamma_{\omega p}=\left(
                                                               \begin{array}{cc}
                                                                 a_{\omega 2} & -c_{\omega 2} \\
                                                                 -c_{\omega 2} & b_{\omega 2} \\
                                                               \end{array}
                                                             \right)$.
  According to (\ref{we7}), the lower bound of the robustness of nonclassicality is
\begin{equation}\label{we22}
  \left[\frac{\det(\gamma_{\omega}+I)}{\det(\gamma_{\omega}+\gamma)}\right]^{\frac{1}{2}}=(f_{1}f_{2})^{\frac{1}{2}},
\end{equation}
with
\begin{equation}\label{we23}
  f_{j}=\frac{(a_{\omega j}+1)(b_{\omega j}+1)-c_{\omega j}^2}{(a_{\omega j}+a)(b_{\omega j}+b)-(c_{\omega j}+c_{j})^2},
\end{equation}
 for $j=1,2$. The strategies of evaluating the supreme of (\ref{we22}) are different for different states.

\subsubsection*{B.1 Two-mode squeezed thermal state}

A two-mode squeezed thermal state is $\rho=S_{2}(r)\tau_{\nu_{1}\nu_{2}}S_{2}^{\dagger}(r)$, where $S_{2}(r)=e^{r(\hat{a}_{1}^{\dagger}\hat{a}_{2}^{\dagger}-\hat{a}_{1}\hat{a}_{2})}$ is the two-mode squeezing operator and $\tau_{\nu_{1}\nu_{2}}$ is the product two mode thermal state with symplectic eigenvalues $\nu_{1},\nu_{2}$, respectively. We have the relationships $a,b=\nu_{+}\cosh(2r)\pm\nu_{-}, c=\nu_{+}\sinh(2r)$, where ${\nu_{\pm}}=\frac{1}{2}(\nu_{1}\pm\nu_{2})$.

For the lower bound of robustness of the state, we choose the CM of witness, $\gamma_{\omega}$, with $a_{\omega 1}=a_{\omega 2}=a_{\omega}, b_{\omega 1}=b_{\omega 2}=b_{\omega}, c_{\omega 1}=c_{\omega 2}=c_{\omega} $ and assume $a_{\omega},b_{\omega}=\kappa_{+}\cosh(2q)\pm\kappa_{-}, c_{\omega}=\kappa_{+}\sinh(2q)$. Then $f_{1}=f_{2}$, we have
\begin{equation}\label{we24}
  R_{\mathcal{C}}(\rho)\geq \lim_{q\rightarrow\infty}f_{1}=\frac{2e^{2r}}{\nu_{1}+\nu_{2}}.
\end{equation}

Notice that (\ref{we24}) is a lower bound which may not be tight for all the two-mode squeezed thermal states. A more general derivation of the lower bound is as follows. Since witness state $\omega$ is supposed to be a Gaussian state, Schr\"{o}dinger-Robertson uncertainty relation $\gamma_{\omega}+i\Delta\geq0$ should be applied. It yields $c_{\omega}\leq \min[\sqrt{(a_{\omega}-1)(b_{\omega}+1)},\sqrt{(a_{\omega}+1)(b_{\omega}-1)}]$. Let $a\geq b$, in order to maximizing $f_{1}$ such that the lower bound is as tight as possible, then the proper choice is $a_{\omega}\geq b_{\omega}$ and it follows $c_{\omega}=\sqrt{(a_{\omega}+1)(b_{\omega}-1)}$. Denote $p=\sqrt{\frac{b_{\omega}-1}{a_{\omega}+1}}$, then
\begin{equation}\label{we25}
  \lim_{a_{\omega}, b_{\omega}\rightarrow\infty}f_{1}^{-1}=\frac{b+1}{2}+\frac{a-1}{2}p^2-cp.
\end{equation}
The minimum of $f_{1}^{-1}$ is achieved at $p=\frac{c}{a-1}$ if there is no further restriction on $p$. However, there is a restriction on $p$ which is $p\in [0,1)$ due to the definition $p=\sqrt{\frac{b_{\omega}-1}{a_{\omega}+1}}$ and $a_{\omega}\geq b_{\omega}$. With the condition $p=\frac{c}{a-1}<1$, namely, $a-c>1$, we have the lower bound $[\frac{b+1}{2}-\frac{c^2}{2(a-1)}]^{-1}$ for robustness. If on the other hand, $a-c<1$ (then $b-c<1$ is also true since we assume $b\leq a$ ), then the minimum of (\ref{we25}) can only be approached at $p=1$, we have the lower bound of robusteness $(a+b-2c)^{-1}=\frac{2e^{2r}}{\nu_{1}+\nu_{2}}$ which is (\ref{we24}). Thus we have the following lower bounds for robustness.
\begin{equation}\label{we26}
  R_{\mathcal{C}}(\rho)\geq\left \{\begin{array}{l}
                              \frac{2e^{2r}}{\nu_{1}+\nu_{2}}, \text{ \quad  \quad  \quad  for } a-c<1, b-c<1,\\
                              \frac{2(a-1)}{(a-1)(b+1)-c^2}, \text{ for } a-c\geq1, b-c<1, \delta<0, \\
                              \frac{2(b-1)}{(a+1)(b-1)-c^2}, \text{ for } a-c<1, b-c\geq1,  \delta<0,\\
                              1 ,   \text{\quad \quad for }  \delta \geq 0
                            \end{array}\right.
\end{equation}
with $\delta=(a-1)(b-1)-c^2$ . Where we include the classical case of $\delta \geq 0$ for completeness.

For the upper bound of robustness, we assume the free state to be $\sigma=S(s)\tau_{\nu_{1}(s)\nu_{2}(s)}S^{\dagger}(s)$, where $\tanh(s)=\sqrt{v_{1s}v_{2s}}$ with $v_{is}=\frac{\nu_{i}(s)-1}{\nu_{i}(s)+1}$. The unnormalized state $\varrho=\tau_{\nu_{1}(s)\nu_{2}(s)}^{-\frac{1}{2}}\rho'\tau_{\nu_{1}(s)\nu_{2}(s)}^{-\frac{1}{2}}\propto \rho''$ is what we are concerned with. The CMs of $\rho'$ and $\rho''$ can be shown to have the same structure as $\rho$ with $a,b,c$ substituted by $a',b',c'$ and $a'',b'',c''$, respectively. We have $a',b'=\nu_{+}\cosh2(r-s)\pm \nu_{-},c'=\nu_{+}\sinh2(r-s)$. A sequence of matrix transformations $\gamma'\rightarrow \beta'\rightarrow \beta''\rightarrow \gamma''$ yield
\begin{equation}\label{we27}
  \left(
    \begin{array}{cc}
      a'' & c'' \\
      c'' & b'' \\
    \end{array}
  \right)+I=2\left(
            \begin{array}{cc}
              1-\frac{1-A}{v_{1s}} & \frac{C}{\sqrt{v_{1s}v_{2s}}}\\
              \frac{C}{\sqrt{v_{1s}v_{2s}}} & 1-\frac{1-B}{v_{2s}} \\
            \end{array}
          \right)^{-1},
\end{equation}
with
$
  \left(
    \begin{array}{cc}
      A & C \\
      C & B \\
    \end{array}
  \right)=2\left(
            \begin{array}{cc}
              a'+1 & c'\\
              c' & b'+1 \\
            \end{array}
          \right)^{-1}.
$

We then introduce an assumption that $\nu_{1}(s)=\nu_{2}(s)=\nu(s)$, hence $v_{1s}=v_{2s}=v_{s}$. With this assumption,
the largest eigenvalue of $\varrho$ turns out to be $\Lambda$, with
\begin{eqnarray}\label{we28}
  \sqrt{\Lambda}=\sqrt{\frac{(1-v_{1})(1-v_{2})}{(1-v)^2}}\frac{2\sinh(s)e^{-\eta}}{1-\tanh(s)}\nonumber\\
  \times[\sqrt{\sinh(2s-\eta-r)\sinh(r-\eta)}\nonumber\\
  +\sqrt{\sinh(2s+\eta-r)\sinh(r+\eta)}]^{-1},
\end{eqnarray}
where $v_{i}=\tanh(\eta_{i}),  \eta_{i}=\frac{1}{2}\log(\nu_{i}),  v=\sqrt{v_{1}v_{2}},  \eta=\frac{1}{2}\log(\frac{1+v}{1-v})$. Then we have
\begin{equation}\label{we29}
  R_{\mathcal{C}}(\rho)\leq \inf_{s}\Lambda=\frac{(1-v_{1})(1-v_{2})}{(1-v)^2}e^{2(r-\eta)}=\frac{2e^{2r}}{\nu_{1}+\nu_{2}},
\end{equation}
with the additional condition that $\rho''$ exists. For the existence of $\rho''$, the determinant of both sides of (\ref{we27}) should be positive. It leads to the condition of
\begin{equation}\label{we30}
  [a'-\nu(s)][(b'-\nu(s)]-c'^2\geq0.
\end{equation}
The inequality can be rewritten as
\begin{equation}\label{we31}
  e^{4s}(1-\nu_{+}e^{-2r})\geq \nu_{+}e^{2r}-\nu_{1}\nu_{2}.
\end{equation}
The minimization of (\ref{we28}) with respect to $s$ leads to a solution $e^{4s}=e^{2r}(e^{2\eta}+e^{-2\eta})-1$. We substitute the solution into the equality of (\ref{we31}). We have
$(1+e^{-2r})e^{2\eta_{1}}-(1-e^{-2r})e^{2\eta_{2}}=2$  and $(1-e^{-2r})e^{2\eta_{1}}-(1+e^{-2r})e^{2\eta_{2}}=-2$ which are $a-c=1$ and $b-c=1$, respectively. The region for the assumption of $\nu_{1s}=\nu_{2s}$ is limited by $a-c<1$ and $b-c<1$, otherwise, we have non-positive determinant of 
the right hand side of (\ref{we27}). 

Outside the region with both conditions $a-c<1$ and $b-c>1$ (or alternatively $a-c>1$ and $b-c<1$), we resort to the free state with $\nu_{1}(s)\neq \nu_{2}(s)$. The upper bound of robustness can be written as
\begin{equation}\label{we32}
  \Lambda=\frac{4}{(1-v_{1s})(1-v_{2s})}\frac{AB-C^2}{(\sqrt{F_{1}}+\sqrt{F_{2}})^2},
\end{equation}
where $F_{1},F_{2}=[1\pm\sqrt{(1-A)(1-B)}/v_{s}]^2-C^2/v_{s}^2$ are functions of $v_{s}$, with $v_{s}=\sqrt{v_{1s}v_{2s}}$. Let $y=\sqrt{\frac{v_{1s}}{v_{2s}}}$, then $\Lambda$ is a function of $y$, but $y$ only appear at the first factor of (\ref{we32}). Notice that the positivity of (\ref{we27}) now is 
\begin{eqnarray}\label{we33}
 [a'-\nu_{1}(s)][b'-\nu_{2}(s)]-c'^2>0,
\end{eqnarray}
 which yields a quadratic inequality $(y-y_{1})(y-y_{2})<0$. Our problem is to maximize $(1-v_{1s})(1-v_{2s})=1+v_{s}^2-(y+1/y)v_{s}$ (such that $\Lambda$ is minimized and the up bound is tighten) with respect to $y$ under the condition of $(y-y_{1})(y-y_{2})<0$. There are three cases: (i) $y_{1}\leq1\leq y_{2}$, we have the solution $y=1$, which is the case of $v_{1s}=v_{2s}$ considered above. (ii)$y_{1}\leq y_{2}<1$, the solution is $y\rightarrow y_{2}$ from below. (iii) $1<y_{1}\leq y_{2}$, the solution is $y\rightarrow y_{1}$ from above. As we have shown that the case (i) corresponds to the region of $a-c<1, b-c<1$. It is evident that cases (ii) and (iii) correspond to the regions $a-c\geq1, b-c<1$ and $a-c<1, b-c\geq1$, respectively. What left is to minimize $\Lambda$ with respect to the single parameter $s$. Numerical calculation shows that the upper bound $\inf_{y,s}\Lambda$ meets with the lower bound in (\ref{we26}) very well. The exact robustness of nonclassicality for two-mode squeezed thermal state is just the right hand side of (\ref{we26}).

\subsubsection*{B.2 Two-mode Gaussian state with standard form of CM}
   The witness operator $\omega$ with 6 parameter CM $\gamma_{\omega}$ is a Gaussian state, thus  Schr\"{o}dinger-Robertson uncertainty relation $\gamma_{\omega}+i\Delta\geq0$ is applied, which yields 
\begin{eqnarray}\label{we34}
&& (a_{\omega 1}b_{\omega 1}-c_{\omega 1}^2)(a_{\omega 2}b_{\omega 2}-c_{\omega 2}^2)\nonumber\\
&&+2c_{\omega 1}c_{\omega 2}-a_{\omega 1}a_{\omega 2}-b_{\omega 1}b_{\omega 2}+1\geq0.
\end{eqnarray}
In the actual calculation of the lower bounds of nonclassicality and entanglement using (\ref{we7}) and (\ref{we13a}), respectively, as $\gamma_{\omega}\rightarrow \infty$, we have 
\begin{eqnarray}
a_{\omega 1}b_{\omega 1}-c^2_{\omega 1}=\epsilon_{1} c_{\omega 1}, \label{we35}\\
a_{\omega 2}b_{\omega 2}-c^2_{\omega 2}=\epsilon_{2} c_{\omega 2}, \label{we36}
\end{eqnarray}
 with positive and finite $\epsilon_{i} (i=1,2)$. The left hand side of (\ref{we34}) is almost nullified and the equality in (\ref{we34}) is nearly achieved in the interative calculation of the lower bound of resources. With assumptions (\ref{we35}) (\ref{we36}) suggested by numerical calculation, we approximately have $(2+\epsilon_{1}\epsilon_{2})c_{\omega 1}c_{\omega 2}\geq a_{\omega 1}a_{\omega 2}+b_{\omega 1}b_{\omega 2}$. Notice that $c_{\omega 1} \approx \sqrt{a_{\omega 1}b_{\omega 1}}$, $c_{\omega 2} \approx \sqrt{a_{\omega 2}b_{\omega 2}}$, let $x=\sqrt{\frac{a_{\omega 1}a_{\omega 2}}{b_{\omega 1}b_{\omega 2}}}$, then we have  $x^2+1-(2+\epsilon_{1}\epsilon_{2})x\leq 0$. The solution 
\begin{equation}\label{we37}
x_{\pm}=1+\frac{1}{2}\epsilon_{1}\epsilon_{2}\pm\frac{1}{2}\sqrt{\epsilon^2_{1}\epsilon^2_{2}+4\epsilon_{1}\epsilon_{2}},
\end{equation}
 is usful in the following calculation.

An especially simple case is $\epsilon_{1}=\epsilon_{2}=0$, hence $x=1$. Thus $a_{\omega1}a_{\omega 2}=b_{\omega1}b_{\omega 2}$. Then for the lower bound of the robustness of nonclassicality, from (\ref{we22}) and (\ref{we23}), we have  
\begin{equation}\label{we38}
R_{\mathcal{C}}\geq \sup_{p}\sqrt{\frac{(1+p^2)(1+p^{-2})}{(b+ap^2-2c_{1}p)(b+ap^{-2}-2c_{2}p^{-1})}}.
\end{equation}
For the lower bound of robustness of entanglement, from (\ref{we13a}), we have
\begin{equation}\label{we39}
R_{\mathcal{E}}\geq \sup_{p}\frac{2}{\sqrt{(b+ap^2-2c_{1}p)(b+ap^{-2}-2c_{2}p^{-1})}}.
\end{equation}
Both lower bounds are analytical and shown in Fig.1 as `lower bound 1', since $p$ is a solution of a power 4 algebric equation in either (\ref{we38}) or (\ref{we39}).  
The `lower bound 1' fits with the upper bound well when the robustness is larger enough (namely, for larger $c_{1}$ and $c_{2}$ in Fig.1), either for robustness of nonclassicality or entanglement. Thus parts of the robustness can be analytically decribed. 

For the non-zero $\epsilon_{1},\epsilon_{2}$, we have the numerical optimized result, denoted as `lower bound 2' in Fig.1. 
\begin{equation}\label{we40}
R_{\mathcal{C}}\geq \sup_{p,\epsilon_{1},\epsilon_{2}}\sqrt{\frac{(1+\epsilon_{1}p+p^2)(1+\epsilon_{2}q+q^{\textbf{}2})}{[b+ap^2+(\epsilon_{1}-2p)c_{1}][b+aq^{2}+(\epsilon_{2}-2q)c_{2}]}},
\end{equation}
and 
\begin{equation}\label{we41}
R_{\mathcal{E}}\geq \sup_{p,\epsilon_{1},\epsilon_{2}}\frac{x_{-}+1+\sqrt{x_{-}\epsilon_{1}\epsilon_{2}}}{\sqrt{[b+ap^2+(\epsilon_{1}-2p)c_{1}][b+aq^{2}+(\epsilon_{2}-2q)c_{2}]}},
\end{equation}
with $q=\frac{x_{-}}{p}$. The condition $q=\frac{x_{-}}{p}$ comes from numerical results. We may calculate the supereme of the right hand side of (\ref{we40}) or (\ref{we41}) with respect to four parameters $p,q, \epsilon_{1},\epsilon_{2}$, then verify $q=\frac{x_{-}}{p}$. Notice that $x=\frac{1}{pq}$ as we have set $p=\sqrt\frac{b_{\omega}}{a_{\omega}}$, $q=\sqrt\frac{b_{\omega 2}}{a_{\omega 2}}$. Thus $pq=x_{-}=\frac{1}{x_{+}}$. It is reasonable that the supreme is achieved when $x$ reaches its edge value $x_{+}$.

For the lower bound of robustness of nonclassicality, there is a case that one of the parameters $\epsilon_{1},\epsilon_{2}$ is no longer finite. Let $\epsilon_{2}\rightarrow\infty$, we have $a_{\omega 2}b_{\omega 2}-c^2_{\omega 2}\propto c^2_{\omega 2}$, thus $f_{2}\rightarrow 1$. Meanwhile we have $\epsilon_{1}\rightarrow 0$ and kept (\ref{we34}) to be true. Using (\ref{we23}), with $\gamma_{\omega}\rightarrow\infty$, we have
\begin{equation}\label{we48}
R_{\mathcal{C}}\geq\sqrt{\frac{2}{a+b-\sqrt{(a-b)^2+4c_{1}^2}}}.
\end{equation}
This lower bound for the robustness of nonclassicality is not shown in Fig.1 (a). The bound is a horizental line if it is displayed and it fits the upper bound well. 

For the upper bounds of the robustnesses of nonclassicality and entanglement, it is not difficult to find the free Gaussian state $\sigma$ since the separability of a two-mode Gaussian state is well known\cite{Duan}\cite{Simon}. The problem of decomposing a Gaussian state to the its symplectic transform and symplectic eigenvalues is also known. The symplectic eigenvalues of a Gaussian state are determined by Williamson theorem\cite{Williamson}. It was only recently that the symplectic transformation was worked out explicitly\cite{Pirandola}. The explict formula is helpful for numerical calculation of the upper bounds of robustness of resources. 

The upper bound of robustness of nonclassicality is calculated in the following way. First we set a six parameter CM $\gamma_{s}$ for the free state $\sigma$, we only set the diagonal elements of the CM, the off-diagonal elements are determined due to the fact that $\sigma$ is on the boundary of classical state set. More concretely, $\gamma_{s}=\gamma_{\sigma x}\oplus\gamma_{\sigma p}$, where 
\begin{equation}\label{we49}
\gamma_{\sigma x}=\left(
            \begin{array}{cc}
              a_{\sigma1} & c_{\sigma1}\\
              c_{\sigma1}& b_{\sigma1} \\
            \end{array}
          \right),
\gamma_{\sigma p}=\left(
            \begin{array}{cc}
              a_{\sigma 2} & -c_{\sigma 2}\\
              -c_{\sigma 2} & b_{\sigma2} \\
            \end{array}
          \right),
\end{equation}
with $c_{\sigma i}=\sqrt{(a_{\sigma i}-1)(b_{\sigma i}-1)}$ for $i=1,2$. The upper bound of robustness of nonclassicality for a two-mode Gaussian state $\rho$ with standard CM then is obtained as the largest eigenvalue of $\sigma^{-\frac{1}{2}} \rho \sigma^{-\frac{1}{2}} $. The upper bound become tight when it is minimized with parameters $a_{\sigma j}, b_{\sigma j}$  for $j=1,2$.

There is a special case that an alternative strategy should be applied. The case occurs when lower bound becomes a horizental line [for small $c_{2}$ and it is not shown in Fig.1(a)]. We use $c_{\sigma 2}=\kappa\sqrt{(a_{\sigma 2}-1)(b_{\sigma 2}-1)}$ instead to achieve the tight upper bound, where $0<\kappa\leq 1$. The results are shown in Fig.1(a), they does not rely on $c_{2}$. The minimization of the upper bound is with respect to five parameters $a_{\sigma j}$, $b_{\sigma j}$  for $j=1,2$ and $\kappa$.

The upper bound of the robustness of entanglement is calculated with the consideration of $\sigma$ being on the boundary of separable state set. That is $\gamma_{\sigma}+i\Delta\geq1$ with the equality being achieved. More concretely, we have $\det(\gamma_{\sigma}+i\Delta)=0$. Thus the upper bound is minimized with respect to five free parameters, see $a_{\sigma j}$, $b_{\sigma j}$  for $j=1,2$ and $c_{\sigma 1}$.

We can see that in all the regions of $c_{2}$ in Fig.1(a) and Fig.1(b), the upper bounds and the best lower bounds coincide with each other. Thus the robustness of nonclassicality and robustness of entanglement are obtained.

\subsection*{C. Multimode symmetric Gaussian state}

\subsubsection*{C.1 Nonclassicality and entanglement}
To describe an $n$ mode symmetric Gaussian states, we use an $n\times n$ matrix function $H_{n}(x,y)$ with all its diagonal elements being $x$, and all off-diagonal elements being $y$. The CM of a symmetric Gaussian state $\rho$ could be $\gamma=H_{n}(a,c_{1})\oplus H_{n}(b,-c_{2})$ with $a>0,b>0,c_{1}\geq0,c_{2}\geq0$ without lose of generality. A symplectic transform $\mathcal{S}=X\oplus X$ will diagonalize $\gamma$, with (for example)
\begin{equation}\label{we50}
X=\left(
            \begin{array}{ccccc}
              \frac{1}{\sqrt{n}} & \frac{1}{\sqrt{n}} & \frac{1}{\sqrt{n}} &...&\frac{1}{\sqrt{n}}\\
              \frac{1}{\sqrt{2}} & -\frac{1}{\sqrt{2}} &0 &...&0\\
              \frac{1}{\sqrt{6}} & \frac{1}{\sqrt{6}} & -\frac{2}{\sqrt{6}} &...&0\\
              ...& ...&...&...&...\\
              \frac{1}{\sqrt{n(n-1)}} & \frac{1}{\sqrt{n(n-1)}} &\frac{1}{\sqrt{n(n-1)}} &...&-\frac{n-1}{\sqrt{n}}\\
            \end{array}
          \right),
\end{equation}
It is clear that $X^{-1}=X^{T}$. Thus $X$ is a orthogonal transform, and $\mathcal{S}$ is an orthogonal symplectic transform. 

For the robustness of entanglment of state $\rho$, consider the fully separable state $\sigma$ with CM, $\gamma_{\sigma}=H_{n}(a_{\sigma},c_{\sigma1})\oplus H_{n}(b_{\sigma},-c_{\sigma2})$. Assuming $a_{\sigma}>0,b_{\sigma}>0,c_{\sigma1}\geq0,c_{\sigma2}\geq0$, the fully separable condition of $\sigma$ is \cite{Chen23PRA}
\begin{equation}\label{we51}
    (a_{\sigma}-c_{\sigma1})[b_{\sigma}-(n-1)c_{\sigma2}]\geq1.
\end{equation}

The eigenvalues of $\sigma^{-\frac{1}{2}}\rho\sigma^{-\frac{1}{2}}$ are not changed by unitary transform. Let $U$ be the unitary transform induced by symplectic transform $\mathcal{S}$. Then $U\rho U^{\dagger}$ has a diagonal CM, $\mathcal{S}\gamma\mathcal{S}^{T}$. Thus $U\rho U^{\dagger}$ is a product of single-mode squeezed states. We denote it as $U\rho U^{\dagger}=\rho_{1}\otimes\rho_{2}^{\otimes(n-1)}$,  where $\rho_{1}$ has a CM of ${\rm diag}[a+(n-1)c_{1},b-(n-1)c_{2}]$, $\rho_{2}$ has a CM of ${\rm diag}(a-c_{1},b+c_{2})$. Similarly, $U\sigma U^{\dagger}$ is a product of single-mode squeezed states. We denote it as $U\sigma U^{\dagger}=\sigma_{1}\otimes\sigma_{2}^{\otimes(n-1)}$. We thus have
\begin{equation}\label{we52}
 U\sigma^{-\frac{1}{2}}\rho\sigma^{-\frac{1}{2}}U^{\dagger}=(\sigma_{1}^{-\frac{1}{2}}\rho_{1}\sigma_{1}^{-\frac{1}{2}})\otimes(\sigma_{2}^{-\frac{1}{2}}\rho_{2}\sigma_{2}^{-\frac{1}{2}})^{\otimes(n-1)}.
 \end{equation}
 The upper bound of the robustness of entanglement is decomposed to single-mode problem. 
 
 For simplicity, we consider the multimode CV-GHZ state in thermal environment. Then we have $a=(2N+1)\frac{1}{n}[e^{2r}+(n-1)e^{-2r}], c_{1}=c_{2}=c=(2N+1)\frac{1}{n}[e^{2r}-e^{-2r}], b=a+(n-2)c$. Further, we donote $2N+1=e^{2\eta}$,  the CMs of $\rho_{1}$ and $\rho_{2}$ are ${\rm diag}(e^{2r+2\eta}, e^{-2r+2\eta})$ and ${\rm diag}(e^{-2r+2\eta}, e^{2r+2\eta})$, respectively.  Denote
 \begin{eqnarray}
 e^{2s_{1}+2r_{1}}=a_{\sigma}+(n-1)c_{\sigma1}, \nonumber\\
 e^{2s_{1}-2r_{1}}= b_{\sigma}-(n-1)c_{\sigma2},\nonumber\\
 e^{2s_{2}-2r_{2}}=a_{\sigma}-c_{\sigma1}, \nonumber\\
 e^{2s_{2}+2r_{2}}= b_{\sigma}+c_{\sigma2}.\nonumber
 \end{eqnarray}
 The CMs of $\sigma_{1}$ and $\sigma_{2}$ are 
${\rm diag}(e^{2s_{1}+2r_{1}},e^{2s_{1}-2r_{1}})$ and  ${\rm diag}(e^{2s_{2}-2r_{2}},e^{2s_{2}+2r_{2}})$, respectively. The equality in the separable condition (\ref{we51}) leads to 
\begin{equation}\label{we53}
s_{1}+s_{2}=r_{1}+r_{2}.
\end{equation}
The upper bound of the robustness of entanglement for state $\rho$ is 
\begin{equation}\label{we54}
\Lambda=\Lambda_{1}\Lambda_{2}^{n-1},
\end{equation} 
with 
\begin{eqnarray}
&&\Lambda_{1}=\frac{2\sinh(s_{1})e^{-\eta}}{1-\tanh(s_{1})}\nonumber\\
&&\times[\sqrt{\sinh(s_{1}+r_{1}-\eta-r)\sinh(s_{1}-r_{1}-\eta+r)}\nonumber\\
&&+\sqrt{\sinh(s_{1}-r_{1}+\eta+r)\sinh(s_{1}+r_{1}+\eta-r)}]^{-1},\nonumber\\
&&\Lambda_{2}=\frac{2\sinh(s_{2})e^{-\eta}}{1-\tanh(s_{2})}\nonumber\\
&&\times[\sqrt{\sinh(s_{2}-r_{2}-\eta+r)\sinh(s_{2}+r_{2}-\eta-r)}\nonumber\\
&&+\sqrt{\sinh(s_{2}+r_{2}+\eta-r)\sinh(s_{2}-r_{2}+\eta+r)}]^{-1},\nonumber
\end{eqnarray}
subject to (\ref{we53}).
With the parameters $s_{1},s_{2},r_{1},r_{2}$ and equality, we can minimize the upper bound of the robustness of entanglement for state $\rho$. The numerical minimization leads to the result of
\begin{eqnarray}
 s_{1}-r_{1}=r-\eta \label{we55}\\
 s_{2}=\eta, \text{\quad}  r_{2}=r. \label{we56}
 \end{eqnarray}
 The condition (\ref{we56}) means $\sigma_{2}=\rho_{2}$ and $\Lambda_{2}=1$. The condition (\ref{we55}) leads to
 \begin{eqnarray}\label{we57}
 \Lambda_{1}=\frac{2\sinh(s_{1})e^{-\eta}}{1-\tanh(s_{1})}[\sqrt{\sinh(2s_{1}-2r)\sinh(2r-2\eta)}\nonumber\\
+\sqrt{\sinh(2r)\sinh(2s_{1}+2\eta-2r)}]^{-1},
 \end{eqnarray}
 A comparison of (\ref{we57}) with (\ref{we18}) yields
\begin{equation}\label{we58}
  R_{\mathcal{E}}(\rho) \leq \inf_{s_{1}}(\Lambda_{1})=e^{2r-2\eta}.
\end{equation}
Hence, for an $n$-mode CV-GHZ state passing through parellel thermal channels, its robustness of entanglement is upper bounded by $e^{2r-2\eta}$, where $r$ describe the CV-GHZ state, $\eta=\frac{1}{2}\log(2N+1)$ describes the thermal channel. The fully separable state $\sigma$ is with a CM of $\gamma_{\sigma}=\mathcal{S}^{T}[{\rm diag}(e^{4s_{1}-2r+2\eta},e^{2r-2\eta})\otimes{\rm diag}(e^{-2r+2\eta},e^{2r+2\eta})^{\otimes n-1}]\mathcal{S}$, where $s_{1}=\frac{1}{4}\log(e^{4r}+e^{4r-4\eta}-1)$.

Next, we consider the lower bound of the robustness of entanglement. Suppose that the conjecture of the maximal mean of a multimode Gaussian  operator over product single-mode states being acheved by product single-mode Gaussian states is true, then we have the tight lower bound for the roustness of entanglement of a multimode Gaussian state. The conjecture has been numerically verified to be true for three-mode symmetric Gaussian operators \cite{ChenPRR}. The lower bound could be 
\begin{equation}\label{we59}
R_{\mathcal{E}}(\rho_{GHZ})\geq e^{2r},
\end{equation}
for a CV-GHZ state of $n$ mode. We will illustrate this lower bound in the following. We consider an $n$ mode symmetric Gaussian state $\rho$ with CM, $\gamma=H_{n}(a,c_{1})\oplus H_{n}(b,-c_{2})$. Let the witness operator $\omega$ be with CM, $\gamma_{\omega}=H_{n}(a_{\omega},c_{\omega1})\oplus H_{n}(b_{\omega},-c_{\omega2})$. The conjecture leads to
\begin{equation}\label{we60}
\sup_{|\psi\rangle=\prod_{j}|\psi_{j}\rangle}\langle\psi |\omega|\psi\rangle=\frac{2^n}{\inf_{x}\sqrt{\det[\gamma_{\omega}+{\rm diag}(xI,\frac{1}{x}I)]}}.
\end{equation}
Denote $e=a+(n-1)c_{1}, f=a-c_{1}, g=b-(n-1)c_{2}, h=b+c_{2}$, $e_{\omega}=a_{\omega}+(n-1)c_{\omega1}, f_{\omega}=a_{\omega}-c_{\omega1}, g_{\omega}=b_{\omega}-(n-1)c_{\omega2}, h_{\omega}=b_{\omega}+c_{\omega2}$. The lower bound for robustness of entanglement will be $\Omega$ according to (\ref{we3}), with
\begin{equation}\label{we61}
\Omega^2=\sup_{\gamma_{\omega}}\inf_{x}\frac{(e_{\omega}+x)(f_{\omega}+x)^{n-1}(g_{\omega}+\frac{1}{x})(h_{\omega}+\frac{1}{x})^{n-1}}{(e_{\omega}+e)(f_{\omega}+f)^{n-1}(g_{\omega}+g)(h_{\omega}+h)^{n-1}}.
\end{equation}
Numerical calculation shows that the extreme is achieved at $e_{\omega}\rightarrow\infty, h_{\omega}\rightarrow\infty, g_{\omega}\approx \frac{1}{e_{\omega}}\rightarrow 0$. Hence 
\begin{equation}\label{we62}
\Omega^2=\sup_{f_{\omega}}\inf_{x}\frac{(f_{\omega}+x)^{n-1}\frac{1}{x}}{(f_{\omega}+f)^{n-1}g}.
\end{equation}
Then it follows $x=\frac{f_{\omega}}{n-2}$, $f_{\omega}=(n-2)f$. At last, we have $\Omega^2=(fg)^{-1}$, hence
\begin{equation}\label{we63}
R_{\mathcal{E}}(\rho)\geq \max\{\frac{1}{\sqrt{(a-c_{1})[b-(n-1)c_{2}]}},1\}
\end{equation}
The lower bound of robustness of entanglement is 
\begin{equation}\label{we64}
R_{\mathcal{E}}(\rho_{GHZ})\geq e^{2r}.
\end{equation}
for a CV-GHZ state.

\subsubsection*{C.2 Numerical verification of the extremality of Gaussian witness}
Consider an $n$-mode symmetric Gaussain witness state with CM $\gamma_{\omega}=H_{n}(a_{\omega},c_{\omega1})\oplus H_{n}(b_{\omega},c_{\omega2})$. The eigenvalues of $H_{n}(x,y)$ are  $x+(n-1)y$ and $(n-1)$ fold of $x-y$, so we may denote the matrix $H_{n}(x,y)$ with another matrix function $F_{n}(x+(n-1)y,x-y)=H_{n}(x,y)$. It follows that the function $F_{n}(\cdot,\cdot)$ has properties of (i), $aF_{n}(x,y)+bF_{n}(u,v)=F_{n}(ax+bu,ay+bv)$ and (ii), $F_{n}(x,y)F_{n}(u,v)=F_{n}(xu,yv)$. To ease the notation, we denote $a=a_{\omega}+(n-1)c_{\omega1}$, $b=a_{\omega}-c_{\omega1}$, $c=b_{\omega}+(n-1)c_{\omega2}$, $d=b_{\omega}-c_{\omega2}$. Then we have $\gamma_{\omega}=F_{n}(a,b)\oplus F_{n}(c,d)$.

Our aim is to write the Gaussian witness in Fock basis, thus the mean of the witness over arbitary product pure states can be evaluated. We need a series of tranformations. Firstly, we apply presqueezing to the witness such that $\gamma_{\omega}\rightarrow\gamma_{\omega}(y)=\frac{1}{y}\gamma_{\omega x}\oplus y\gamma_{\omega p}$. Then we transform it to complex CM  $\tilde{\gamma}_{\omega}(y)$ and further to 
$\beta(y)=(\sigma_{3}\otimes I_{n})[\frac{1}{2}(\tilde{\gamma}_{\omega}(y)+\sigma_{1}\otimes I_{n})]^{-1} (\sigma_{3}\otimes I_{n})$ as we have done in the main text.  At last, we have 
\begin{equation}\label{we65}
\sigma_{1}\otimes I_{n}+\beta(y)=\left( \begin{array}{cc}
   F_{n}(a'(y),b'(y)) & F_{n}(c'(y),d'(y)) \\
  F_{n}(c'(y),d'(y))& F_{n}(a'(y),b'(y)) \\
 \end{array}
\right)
\end{equation}
with $a'(y)=\frac{a/y-cy}{(a+y)(c+1/y)},b'(y)=\frac{b/y-dy}{(b+y)(d+1/y)},c'(y)=\frac{ac-1}{(a+y)(c+1/y)},d'(y)=\frac{bd-1}{(b+y)(d+1/y)}$.  We choose $a'(y)+(n-1)b'(y)=0$ to nullify the the diagnal elements of the matrix $\sigma_{1}\otimes I_{n}+\beta(y)$, it will greatly simplify the evaluation of the mean of Gaussian witness in Fock basis. It leads to 
\begin{equation}\label{we66}
  -\frac{y}{a+y}+\frac{1}{cy+1}-\frac{(n-1)y}{b+y}+\frac{n-1}{dy+1}=0.
\end{equation}
The solution of (\ref{we66}) yields a presqueezed Gaussian witness $\omega(y)$ with nullified diagonal elements of its matrix $\sigma_{1}\otimes I_{n}+\beta(y)$. Then in Fock basis, the matrix element of $\omega(y)$ is 
\begin{equation}\label{we67}
  \omega(y)_{l,m}=\sqrt{\beta(y)}\mathcal{O}_{l,m}(t,t')e^{\frac{1}{2}(t,t')(\sigma_{1}\otimes I+\beta(y))(t,t')^T}.
 \end{equation}
Let the product state be $|\psi\rangle=\Pi_{j=1}^{n}|\psi_{j}\rangle$, with $|\psi_{j}\rangle=\sum_{k_{j}}c_{k_{j}}|k_{j}\rangle$. Denote $|k\rangle=|k_{1}k_{2},...,k_{n}\rangle$, then mean of the witness over the product state is $\langle\psi|\omega(y)|\psi\rangle=\sqrt{\det\beta(y)}M_{0}$, with
\begin{equation}\label{we68}
M_{0}=\sum_{l,m}c^{*}_{l}c_{m}\mathcal{O}_{l,m}(t,t')e^{\frac{1}{2}(t,t')(\sigma_{1}\otimes I+\beta(y))(t,t')^T},
\end{equation}
where $c_{m}=\Pi_{j=1}^{n}c_{m_{j}}$.
The iterative numerical calculation shows that the maximum is achieved at product vaccum state $|\psi\rangle=|0^{\otimes n}\rangle$. The maximal value of $M_{0}$ is $1$. In the numerical calculation, we set a cutoff $k_{\rm cut}$ for the Fock basis $|k\rangle=|k_{1}k_{2},...,k_{n}\rangle$ such that $k_{j}<k_{\rm cut}$ for all $j$. The comupational overhead for building a presqueezed three mode witness in Fock basis is propotional to $k_{\rm cut}^{15}$, it is about half an hour when $k_{\rm cut}=5$ and about a minite when $k_{\rm cut}=4$. The comupational overhead for iteration and summation is almost neglectible.  

One the other hand, maximizing $\det\beta(y)$ with respect to $y$ also yields the equation (\ref{we66}). 

\end{document}